# Gauging functional brain activity: from distinguishability to accessibility

David Papo*

*SCALab, CNRS, Université de Lille, Villeneuve d'Ascq, France*
*Correspondence:papodav@gmail.com*

ARTICLE INFO

ABSTRACT

Standard neuroimaging techniques provide non-invasive access not only to human brain anatomy but also to its physiology. The activity recorded with these techniques is generally called *functional* imaging, but what is observed per se is an instance of *dynamics*, from which functional brain activity should be extracted. Distinguishing between bare dynamics and genuine function is a highly non-trivial task, but a crucially important one when comparing experimental observations and interpreting their significance. Here we illustrate how neuroimaging's ability to extract genuine functional brain activity is bounded by functional representations' *structure*. To do so, we first provide a simple definition of functional brain activity from a system-level brain imaging perspective. We then review how the properties of the space on which brain activity is represented allow defining relations ranging from distinguishability to accessibility of observed imaging data. We show how these properties result from the structure defined on dynamical data and dynamics-to-function projections, and consider some implications that the way and extent to which these are defined have for the interpretation of experimental data from standard system-level brain recording techniques.

## INTRODUCTION

System-level neuroimaging techniques such as PET and MRI make it possible to noninvasively access not only the anatomy of the human brain but also its physiology (Raichle, 2000). Brain activity recording with these techniques is generally called *functional* imaging, the term being equally applied to electrophysiological techniques such as EEG or MEG. However, observed activity is not genuinely functional *per se*, and neuroimaging data should a priori be treated as brain *dynamics*. Extracting functional brain activity from bare dynamics represents a non-trivial though often implicit process (Atmanspacher and beim Graben, 2007; Allefeld et al., 2009).

It is often important to compare representations associated with different recording sessions from the same individual, different individuals, or experimental conditions. Comparisons can come in various forms, which can be expressed in terms of the question they address: when are two representations distinguishable? How far apart are they? What do neighbouring representations look like? Is a transition possible from a given representation to another?

Here we illustrate how neuroimaging's ability to address these questions is bounded by functional representations' *structure*. To do so, we first provide a simple but convenient definition of functional brain activity from a system-level brain imaging perspective, a more comprehensive definition being beyond the scope of the present work. We then review how the properties of the space on which brain activity is represented allow defining properties ranging from distinguishability to accessibility. We show how these properties result from the structure defined on dynamical data and dynamics-to-function projections, and consider some implications that the way and extent to which these are defined have for the interpretation of experimental data from standard system-level brain recording techniques.

## DEFINING FUNCTIONAL BRAIN ACTIVITY

It is logical to understand *functional* activity as one associated with some *function*. Function can be defined as the ability to perform a given cognitive or physiological task, imposed by the environment. Insofar as individuals' behavioural performance can be thought of as resulting from brain properties, functional activity refers to both behaviour and neural structures. This double use reflects two complementary goals, i.e. understanding how brain anatomical structure and dynamics unfolding on it control function, and how performance of cognitive or physiological tasks acts on brain anatomy and dynamics, producing functional subdivisions in the brain. In the former, a space $\Psi$ of (typically non-observable) cognitive or physiological functions $\{\psi_1, \psi_2, ..., \psi_J\}$ is described using a finite set $\{\varphi_1, \varphi_2, ..., \varphi_K\} \in \Phi_{Obs}$ of carefully selected coarse-grained aspects of brain anatomy or physiology (reflecting at a macroscopic level neurophysiological phenomena $\Phi_{NObs}$ not observable when using a given system-level neuroimaging technique) associated with observable fitness or performance measures $\{\gamma_1, \gamma_2, ..., \gamma_L\} \in \Gamma$ from subjects at rest or carrying out given tasks. In the latter, the ability to carry out given tasks is used as a probe exposing information on brain properties $\Phi$. Each of these levels is in general endowed with some structure $S$, i.e. some relationship among its elements. Thus, defining functional brain activity using system-level neuroimaging techniques involves partitioning two complex spaces, respectively made observable by behaviour and brain recording techniques, putting a structure on the set of equivalence classes, and mapping the corresponding structures.

BRAIN PARCELLATION

Characterizing functional activity is in essence a parcellation problem. When using $\Phi$ to make sense of $\Psi$, one ultimately aims at partitioning the space of cognitive functions $f: \Psi \to \Psi/\mathcal{R}$, where $\Psi = \Psi(\Gamma, \Phi)$, and $\Psi/\mathcal{R}$ is the space of equivalence classes under the relation $\mathcal{R}$. In the opposite case, $\Phi$ is partitioned into functionally meaningful units $g: \Phi \to \Phi/\mathcal{R}'$ using cognitive tasks as probes. This implies evaluating the sets $U = \pi^{-1}(V)$, where $U \subset \Psi, V \subset \Phi/\mathcal{R}', \pi: \Psi \to \Phi/\mathcal{R}'$ and $\mathcal{R}'$ is a relation defined on $\Phi$, or the equivalent in the opposite case. Since typically $L \ll K$, the structure on $\Phi$ is finer than that on $\Gamma$, and physiology is more often used to define the cognitive space than the opposite case. Meaningful functional units correspond to the family of sets $\mathcal{U}_\pi = \{U = \pi^{-1}(V)\}$ (or, equivalently, $\mathcal{V}_{\pi'} = \{V = \pi'^{-1}(U)\}$). How to construct $\mathcal{U}_\pi$ (or $\mathcal{V}_{\pi'}$), what form the corresponding space may take, and therefore what may be regarded as functional, depends on the way $(\Psi, \mathcal{S}_\Psi)$ and $(\Phi, \mathcal{S}_\Phi)$, where $\mathcal{S}$ denotes a generic structure, are defined and mapped onto each other through $\pi$ (or $\pi'$).

Classical neuropsychological descriptions map observed behaviour $\Gamma$ onto the anatomical Euclidean space $\mathfrak{E}$, with orthonormal coordinates, so that $\Phi_\mathfrak{E} \subset \mathbb{R}^3$. Brain lesions induce a coarse partition $\Phi_{\mathfrak{E}/L}.\mathcal{U}_\pi$ is extrapolated from the overlap between lesions and *cortical areas*, i.e. partitions defined on the anatomical space on the basis of cytoarchitecture, histological structure or organization homogeneity (Brodmann, 1909), associated with a map $\Psi \to \Phi_\mathfrak{E}$, using double dissociations in conjunction with the assumption of modularity of both $\Phi_\mathfrak{E}$ and $\Psi$ (Dunn and Kirsner, 2003).

System-level neuroimaging maps $\Psi$ onto some recording technique-specific function of macroscopic observables $\varphi_i \in \Phi$ of brain physiology. Neuroimaging data are typically treated as (scalar, vector or tensor) fields $\mathcal{F} = \{f_X(\vec{s}, t)\}$, where $\vec{s}$ lives in a subspace isomorphic to $\mathbb{R}^3$ and $t \in \mathbb{R}$ is the physical time, and described in terms of some convenient function of this field, in the spatial (anatomical), temporal, frequency domains or in the phase space (either of observed activity, considered as a dynamical system, at experimental time scales, or as a morphospace, e.g. at developmental or evolutionary time-scales). On the other hand, while $\Gamma$ is typically a scalar or vector field, it can sometimes take the form a complex *function space*.

Functional parcellations are defined in a recording technique- and scale-dependent manner. For fast sensory processes, functional equivalence classes can be defined by characterizing the dynamical *range*, i.e. the range of stimulus intensities resulting in distinguishable neural responses, while the dynamical *repertoire*, i.e. the number of distinguishable responses, quantifies the functional phase space extension. How, viz. on what space and along what properties or relations to define distinguishability represents the most crucial questions. On the other hand, for processes such as thinking or reasoning, which lack a characteristic duration and trivial behavioural scales (Papo, 2015) the definition of functional but also dynamical partitions is less straightforward. For instance, $\Phi$ can be locally partitioned into quasi-stable spatiotemporal microstates (Lehman et al., 1987; Kaplan et al., 2005; Allegrini et al., 2010 ; van de Ville et al., 2010; Hadriche et al., 2013; Watanabe et al., 2013), or into predictively equivalent sets induced by the causal states of the dynamics, with equivalence relations grouping all histories that give rise to the same prediction (Crutchfield et al., 2009); or, if one considers relationships between dynamics at different scales, parcellations may be *universality classes*, i.e. regions of the phase space sharing a single scale invariant limit under a renormalization group flow (Papo, 2014). In these latter cases, extracting function from observed dynamics is conceptually and technically arduous and involves understanding the structure of brain dynamics and how this can be used to ultimately define function.

Superstructure of brain imaging representations

The space on which parcellations are defined is in general endowed with some superstructure. First, brain anatomy and dynamics can be endowed with a network structure (Bullmore and Sporns, 2009), and, as a consequence, with topological properties (Boccaletti et al., 2006) and symmetries (Pecora et al., 2014). These induce multiple possible parcellations of the anatomical and dynamical spaces. Community structures constitute a particularly illustrative example of some important aspects worth mentioning in this context. First, community structures can come in qualitatively different forms (e.g. static or dynamic) and can be quantified in various different ways, corresponding to qualitatively different purposes (Schaub et al., 2018). For instance, fuzzy (Simas and Rocha 2015) or overlapping (Palla et al., 2005) community structures may be more appropriate than *stricto sensu* partitions of the space. Second, community structures can take a generalized form, as it is always possible to generate ensembles of networks with given properties inducing qualitatively different underlying spaces (Newman and Peixoto, 2015). Third, community structure and, more generally, network structures indicate that parcellations may not necessarily be local in the anatomical space.

$\Phi$ is typically embedded into the same anatomical Euclidean space $(\mathfrak{E}; d)$ of classical neuropsychology, where $d$ is the usual metric. In the simplest case, it is *de facto* treated as a field $(\mathcal{F}; d)$. This translates the fact that, at least at the temporal scales at which anatomy represents a genuine boundary condition for brain anatomy and ultimately for the dynamics (Papo, 2017), the brain can be thought of as a spatial network (Barthélemy, 2011), reflecting the existence of geometric alongside topological constraints on the overall organization of brain anatomy and dynamics (Robinson, 2013b; Stiso and Bassett, 2018).

While the anatomically-embedded dynamical space as a whole should not be regarded as homeomorphic to the Euclidean space $\mathbb{R}^n$, it may be treated as almost everywhere locally isomorphic to it. One may then represent it as a *topological manifold* $(X, \mathscr{C})$ i.e. as a *paracompact* topological space $X$ equipped with an *atlas*, a cover $\mathscr{C}$ of open sets in which each $C \in \mathscr{C}$ is homeomorphic to an open subset $D \subseteq \mathbb{R}^n$ through a map $\varphi_C: C \to D$ called a *chart* of $\mathscr{C}$ (Robinson, 2013). Finally, whenever brain imaging data can effectively be treated as the output of a dynamical system, they may be modelled as a *topological dynamical system*, i.e. a triple $(\Phi, \mathcal{S}, T_s)$, where $\Phi$ is a Hausdorff (separable) topological space, $\mathcal{S}$ a topological semigroup prescribing the matching conditions between overlapping local trivialization charts, and $T_S$ a continuous function $T_S: \mathcal{S} \times \Phi \to \Phi$. For instance, at long time scales, brain fluctuations are characterized by non-trivial scaling properties such as scale-invariance (Papo, 2013). The set of associated renormalization operators has a multiplicative semigroup structure on the time-scale space and a covariance property (Papo, 2014).

Although a global Euclidean geometry may not be an appropriate description of global dynamics on the neurophysiological space, there is more than one way in which $\Phi$ can nonetheless be equipped with a geometry. First, geometry may be derived from topology. Not only can a network always be embedded in a surface, provided it is of sufficiently high genus (roughly meaning it has a sufficient number of surface handles) (Aste et al., 2005), but geometry may also turn out to be an emergent property of the underlying topology (Bianconi and Rahmede, 2017). Furthermore, time series may be mapped into geometry (Amari and Nagaoka, 2007; Lesne, 2014; Ali et al., 2018). This can be done by representing observed brain activity in terms of probability distribution functions. This induces a smooth manifold $\mathcal{M}$ whose points are probability distributions defined on a common probability space (Amari and Nagaoka, 2007). Working with a probability distribution space allows recovering a continuous space even when the underlying state space is discrete, and representing brain activity as a geometrical structure. Fluctuations' scaling properties may help equipping the space with a specific geometry. For instance, scale-free distributions suggest a *fractal* geometry, while accounting for the history-dependence of brain fluctuations may require a *non-commutative* one or a quasi-metric space.

### GAUGING IMAGING DATA

Interpreting neuroimaging data requires introducing relations among experimental conditions and this, in turn, understanding the implications that given structures have on the definition of the families $\mathcal{U}_\pi$ or, equivalently, $\mathcal{V}_{\pi'}$. Conversely, any interpretation introduces further structure which allows measuring quantities over the considered spaces. On the one hand, endowing data spaces with given structural properties induces specific equivalence classes. For instance, two dynamical systems are dynamically equivalent if they are *topologically conjugate* (Pasemann, 2002). More generally, observed data may be classified up to a given property (e.g. homotopy, conjugacy, symmetry, etc.) or according to *obstructions* to one of them. Conversely, comparing experimental conditions requires comparing their associated (e.g. network) structure, each structure involving its own set of operations restrictions, and sometimes adding further structure (Simas et al., 2015; Schieber et al., 2017; Gadyiaram et al., 2017).

At the most basic level, when comparing experimental conditions, one needs to evaluate the *topological distinguishability* of two sets $V_1$ and $V_2$ in $\Phi/\mathfrak{R}'$ and the corresponding $U_1$ and $U_2$ in $\Psi/\mathfrak{R}$. If one considers the basic field representation of neuroimaging data, this amounts to comparing two fields $f_X$ and $f_Y$, a seemingly rather tractable task. Because the functional space is not everywhere isomorphic to the anatomical one at macroscopic scales, gauging distinguishability in terms of pattern similarity in the anatomical space $(\mathfrak{E}; d)$ can be misleading. Typical representations isomorphic to the anatomical space become difficult to compare when the subspace of $\Phi$ has complex, e.g. non compact geometry. A number of factors, including noise and inter-individual differences, can render the distinguishability of Euclidean representations problematic. No less importantly, functional brain activity may be organized in patterns with similar meaning but considerably different underlying anatomical structure (Ganmor et al., 2015).

The extent to which parcellations can be distinguished from each other is related to the *separation* properties on the space, which in turn determine the possible nature of functions and operations on the space (Dodson and Parker, 1997). The functional space is not necessarily a Hausdorff space, even when $\Phi$ is embedded in the cortical Euclidean space $\mathfrak{E}$. This is clearly the case for fuzzy relations (Grzegorzewski, 2017) or overlapping communities (Palla et al., 2005). When neuroimaging data are endowed with a topological manifold representation, this situation is reflected by overlaps between charts of the manifold's atlas, and *transition functions* are needed to resolve these areas.

Once endowed with some structure, membership of a given equivalence class can be assessed using maximum entropy methods, and observed data are considered as an instance of an ensemble of objects with these properties (Bianconi, 2007; Cimini et al., 2019). The functional significance of given representation can also be gauged by its ability to perform a given task, e.g. classification or prediction, (Zanin et al., 2016).

Often, it is also necessary to quantify how far $V_1$ and $V_2$ and the corresponding $U_1$ and $U_2$ are from each other. This implies defining some property intuitively translating the concept of *distance*. While the anatomically-embedded functional space can only locally be considered a Euclidean metric space, distances may be defined for other structures in a way that is dictated by the structure itself (Rossi et al., 2015; De Domenico and Biamonte, 2016). When operating in a probability distribution space, $\Phi$ can be equipped with the Fisher information metric e.g. by using the covariance matrix as a metric tensor (Crooks, 2007). This endows the space $\mathcal{M}$ of probability distributions with a Riemannian differential manifold structure $(\mathcal{M}, g, \theta)$, where $g$ is the Fisher-Rao information metric and the parameters $\theta$ are probability measures giving the coordinates on this manifold. The Fisher metric converts parameter space distance into a unique measure of distinguishability between models (Amari and Nagaoka, 2007), and can be used to quantify the informational difference between measurements as well as the sensitivity of model predictions to changes in parameters (Machta et al., 2013). On the other hand, whenever neuroimaging data can be treated as a dynamical system, a dynamical distance can be derived from the dynamics itself. This distance allows a coarse graining that is in some sense optimal with respect to the dynamics (Gaveau and Schulman, 2005).

Finally, irrespective of whether one is considering a static (e.g. network) structure or a dynamical system with an explicit dynamical rule, a metric can be induced using perturbation methods. Importantly, these methods can also help defining *proximity* relations in $\Phi$ (Peters, 2016, 2018).

### FROM DYNAMICS TO FUNCTION

To move from *dynamical* equivalence classes, comprising identical dynamical properties and specific phase and parameter space symmetries, to *functional* equivalence classes, comprising patterns of neural activity that can achieve given functional properties (Ma et al., 2009) requires lifting $T_S$ to $\Phi/\Gamma$, and considering the structure induced by the dynamical system $T_{S_\Psi}: \Phi/\Gamma \to \Phi/\Gamma$. While dynamical properties may sometimes readily be interpreted in functional terms, e.g. the co-existence of different attractors points to multi-functionality of a given module or system (Pasemann, 2002), the dynamics-to-function transition may give rise to non-trivial properties that cannot be anticipated based on the dynamics alone. Important properties such a robustness may drastically change when moving from $T_S$ to $T_{S_\Psi}$.

While each recording technique's precision induces specific *a priori* parcellations of $\Phi$, these are in general not functionally relevant. To be functionally meaningful, metrics in $\Phi$ need to be appraised in the space $\Psi$ made observable through $\Gamma$. To do so involves determining how properties in one structure are transferred onto those of the other. This is where the fundamental role of the map $\pi$ (or $\pi'$) comes into play. Ideally, one seeks the finest topology in $\Phi/\mathcal{H}$ that renders the $\pi: \Gamma \to \Phi$ surjection continuous. This means endowing $\Phi$ with the *quotient topology* with respect to $\pi$, i.e. the family $\mathcal{U}_\pi = \{V \mid \pi^{-1}(V) \text{ is open in } \Gamma\}$. In other words, from a neuroimaging view-point, functional brain activity can be thought of as a *fibre bundle*, i.e. a quadruple $(\Psi, \Phi, \pi, \mathcal{U}_\pi)$, where $\Psi$ is the *total space*, $\Phi$ the base $\pi: \Psi \to \Phi$ a continuous surjective function called the *fibre bundle projection*, and $\mathcal{U}_\pi$ the fibres. $\Phi$ can ultimately be identified with a subspace of $\Psi$ through a *section* of the fibre bundle, i.e. a continuous right inverse of the projection function $\pi$ defined on an open set of $\Phi$. Globally, the space $\Psi$ is not necessarily isomorphic to a Cartesian product $\Phi \times \mathcal{U}_\pi$ but may result from gluing together several Cartesian products defined on local open domains of $\Phi$. Transitions between the fibre bundle's local trivializations may lie in a topological group, known as the *structure group*, acting on the fibres $\mathcal{U}_\pi$. Moreover, $\Phi$ is often implicitly assumed to be a *measurable* space. Measurability of the physiological space is used to induce measurability in the cognitive one; a $\sigma$-algebra on $\Psi$ would then precisely be given by the counter-images of elements of the $\sigma$-algebra on $\Phi$ induced by its topology. It is intuitive to understand $\pi$ as a *homomorphism*, i.e. a structure-preserving map between algebraic structures, whenever $\Gamma$ can be equipped with a quasi-metric structure, as in spatial perception or sensory processes, and $\Phi$ is organized in topographic maps, so that neighbours in $\Gamma$ have corresponding ones in the anatomical space $\Phi$. But what are in practice $\pi$'s properties? Is it invertible, continuous, measurable, etc.? Is it an isometry?

Before examining the $\pi: \Psi \to \Phi$ map, it is first worth recalling that system-level neuroimaging representations $\Phi_{Obs}$ are in essence kinetic models averaging over $\Phi_{NObs}$ stochasticities (Zaslavsky, 2002), i.e. there exists a non-observable map $\tilde{\pi}: \Phi_{NObs} \to \Phi_{Obs}$. The space induced by this map can show permutation symmetry but also combinatorial complexity with respect to more fine-grained $\Phi_{NObs}$ configurations (Brezina, 2010). An adequate coarse-graining of $\Phi_{NObs}$ should preserve a microscopic state description exhibiting a given property, e.g. a symmetry (Cross and Gilmore, 2010), and the possibility to obtain a dynamical rule for the system (Allefeld et al., 2009). A faithful representation of the hidden structure at microscopic scales would require finding a generating partition, but this is in practice an excessively arduous task (Kantz and Schreiber, 2004). While macroscopic scale descriptions are *stricto sensu* dynamically emergent states only if they correspond to a Markov coarse-graining of lower-level dynamics (Adler, 1998; Shalizi and Moore, 2003; Bollt and Skufca, 2005; Gaveau and Schulman, 2005; Allefeld et al., 2009), both $\Phi_{Obs}$ and $\Gamma$ can loosely be thought to emerge from the renormalization of neural activity at microscopic scales, i.e. at scales not observable with a given neuroimaging technique. The way microscopic scales renormalize into macroscopic ones can determine the scale at which the space is *locally isomorphic* to $\mathbb{R}^n$ and can effectively be treated as a topological manifold. This scale may be induced by permutation symmetry with respect to a given property at microscopic scales. Thus, macroscopic parcellations in the anatomical space may consist of topographical regions for which such symmetry holds. On the other hand, whether and to what extent $\Phi_{Obs}$'s topological properties actually reflect those of $\Phi_{NObs}$ is not always entirely clear. For example, the robust computational properties associated with motifs in microcircuits (Klemm and Bornholdt, 2005; Gollo and Breakspear, 2014) do not necessarily characterize structurally isomorphic circuits at macroscopic scales.

The $\Phi_\mathfrak{E} \to \Psi$ map associated with the lesion-based framework is in general ill-defined. This is due to the fuzzy lesion contour geometry but, more importantly, 1) to the fact that the functional space is only locally a metric one, 2) to $\Phi_\mathfrak{E}$'s lack of temporal dimension, so that the resulting $\mathcal{U}_\pi$ is too low-dimensional to capture the complexity of both $\Phi$ and $\Psi$; and 3) to the brain's *degeneracy*. In a degenerate system, structurally different elements give rise to the same output. A localized lesion in a degenerate system may produce no cognitive deficit, and it is impossible to establish if a lesioned brain region is a necessary part of a system subserving a given cognitive process (Price and Friston, 2002).

A well-behaved $\Phi \to \Psi$ map can sometimes be obtained. A notable example is represented by Kelso's bimanual motor coordination paradigm, which requires coordinating finger movements and following a pacing metronome whose oscillation frequency is systematically increased (Kelso, 1995). Once the relative phase $\phi$ between the fingers is chosen as the order parameter describing the dynamics, $\Phi$ and $\Psi$ are both differentiable and $\Psi$ turns out to be diffeomorphic to the macroscopic velocity field $\nabla \cdot \Phi$, which in turn can be thought of as collective modes of underlying neurophysiological activity (Kelso et al., 1998). In this special case, the functional space can be considered as a *smooth fibre bundle*, as total space, base and fibre are all smooth manifolds and $\pi$ is surjective. However, in most contexts, it is hard to describe objects in $\Psi$ in terms of differential equations or even more general dynamical rules, and the $\Psi \to \Phi$ map is non-trivial.

While a smooth manifold representation of $\Phi$ may not always be available, relatively well-behaved mappings may occur in other contexts as well. For instance, brain networks display generic hierarchical structure (Meunier et al., 2010), a structure that may be mirrored both by brain dynamics' temporal fluctuations and by a corresponding hierarchical one in $\Psi$. An example may be represented by linguistic functions may be defined in terms of hierarchical relations, rules and operations.

### From measure to accessibility

In a sense, proximity, neighbourhood and distances *lato sensu* are usually quantified in terms of static representations, both for truly quasi-static data as fMRI images, and for dynamic ones such as EEG recordings. However, all these properties actually depend on the way one state in $\Phi/\Gamma$ may be transformed into another under some neurophysiological process.

To understand how the properties of the dynamical space are inherited by the functional space, one may think of neurophysiological processes being only partially observable at the system level of non-invasive neuroimaging techniques, as *genotype*, and of observed behaviour or macroscopic brain activity as the corresponding *phenotype*, resulting from coarse-graining of physiological processes. The crucial question is: what space does the genotype-to-phenotype map induce?

A smooth genotype-to-phenotype map can sometimes be ensured. For instance, in Kelso's paradigm (Kelso et al., 1998), possible functional discontinuities in $\Psi$ can be explained in terms of genuine brain dynamics, as brain dynamics has the structure of a *differentiable manifold*, i.e. a topological manifold with an atlas whose transition functions are differentiable, allowing for differential calculus on the manifold. Function is then defined in terms of dynamical variables, i.e. synchronization and syncopation; components and collective variables in $\Psi$ can both be endowed with explicit differentiable analytical expressions, and cognitive demands can be construed as their boundary conditions (Kelso, 1995).

However, the structure induced by the $\Psi \to \Phi$ map may be highly non-trivial (Stadler et al., 2001): not only may the phenotype space induced by the dynamical system $T_{S_\Psi}: \Phi/\Gamma \to \Phi/\Gamma$ be a non-metric one, but even the less stringent notion of topology may need to be relaxed (Stadler et al., 2001; Stadler, 2006). This can be understood in terms of *accessibility* in the genotype space and of the genotype-to-phenotype map. Whether defined on $\Psi$ or $\Phi$, a robust dynamical notion of nearness and neighbourhood in the phenotype space should reflect the genotype space's accessibility structure, i.e. the variation operators establishing which configurations are accessible from given ones should reflect the dynamics of physiological processes. Ultimately, to define accessibility requires understanding which phenotypic variations are neurophysiologically neutral and which ones are realizable in the neighbourhood of underlying neuronal variations. Since accessibility lacks symmetry in general, nearness in the induced space would be non-symmetric. Furthermore, important dynamical patterns of observed activity, including intermittency, degeneracy, redundancy, and robustness could be dynamic manifestations of the phenotype's topological properties induced by the genotype-to-phenotype map. Ultimately functional complexity is driven by the robustness induced by neutrality in the genotype (Doyle and Csete, 2011). This clarifies the distinction between structural, dynamical and functional robustness (Lesne, 2008).

## Concluding remarks

Defining functional brain activity and ultimately how the brain implements given functions are arduous tasks. Here we did not address this general ontological issue, nor did we address the comparably complex issue of (state space) reconstruction from data, but a rather more circumscribed methodological question: how does the structure that available neuroimaging are endowed with determine our ability to extract genuine functional brain activity?

What is observed of the brain *per se* without a definition of structure is an instance of dynamics or whatever is used to describe $\Phi$. For example, so-called *functional* networks, constructed based on some dynamical coupling metric between brain regions (Bullmore and Sporns 2009) should in general more appropriately be referred to as *dynamical*. Genuine *functional* networks can instead be thought of as partitions or macrostates (Shalizi and Moore, 2003; Allefeld et al., 2009; Wolpert et al., 2014) of dynamical ones, and functional activity as a process unfolding on the network (Papo et al., 2014b). Characterizing functional equivalence classes requires an appropriate auxiliary space, whose properties are those of the superstructure with which observed data are equipped. Likewise, observed activity can be understood as functional or purely dynamical in a given variable in a scale- and technique-dependent manner. Thus, while it is still unclear what neurophysiological function should be ascribed to it in general, spontaneous brain activity should be treated as a dynamical rather than a functional phenomenon when observed at time-scales much shorter than those of its generic properties (Papo, 2013).

Though not necessarily the case, some structures used in data analysis, viz. in topological data analysis, may reflect the way function impacts on brain dynamics. The brain can be thought of as a "geometric engine", implementing structures (e.g. fibre bundles) through task-specific *functional architectures*, hard-wired anatomical apparatus together with some dynamics (Koenderink and van Dorn, 1987; Moser and Moser, 2008), with non-random topological properties (Curto et al., 2017; Giusti et al., 2015) and (possibly non-Euclidean) geometry (Petitot, 2013, 2017). Moreover, representing brain data as probability distributions allows characterizing function as a perturbation of the functional form of brain dynamics, amplitude or frequency modulations representing a short temporal scales special case (Papo, 2014). This representation induces a space of functions on $\Phi$ straightforwardly mirroring the of non-observable space $\Psi$. Furthermore, such a representation would naturally come with a metric and would allow estimating important aspects such as the cost of moving from one probability distribution function to another (Kempes et al., 2017).

Topological signal processing tools are consistent with information locality in some domain (space, time, frequency, etc.) (Robinson, 2013a). On the other hand, functional neuroimaging often boils down to a dynamic way of addressing the definition of cortical area. In spite of novel preparations, imaging and optical methods, quantitative architectonics, and high-performance computing the criteria for defining a cortical area are typically based on structure-function associations at the level of cortical areas in the anatomical space (Amunts and Zilles, 2015). The underlying assumption is that information is (locally) compact in the anatomical support. However, the brain may show genuine non-locality, i.e. interaction-induced emergence as opposed to bare anatomical connectivity, and the role of non-local information cannot be neglected (Santos et al., 2018).

Embedding into a metric space may overshadow important aspects of its structure (Vaccarino et al., 2015); it may then be useful to represent $\Phi$ as a space that does not derive its topology from a metric, and to resort to representations that are non-local in $(\mathfrak{E}; d)$ (Baruchi and Ben-Jacob, 2004; Petri et al., 2014). Such spaces allow treating the multiplicity of observables, observation scales and possibly geometries and defining relationships between geometric objects constructed using different parameter values, and continuous maps relating these objects (Carlsson, 2009).

Endowing $\Phi$ with a structure involves discretionary choices at various levels somehow associated with assumptions on what should be regarded as functional in brain activity, introducing circularity between definition and quantification of functional brain activity (Papo et al., 2014b). For instance, there are no criteria to elect the space to be equipped with a network structure, and to define its boundaries, constituent nodes and edges (Papo et al., 2014a). Brain function "stylized facts", topological and geometrical concepts, and thorough behavioural studies (Krakauer et al., 2017) may all help defining and quantifying brain function. Finally, functional representations should highlight neurophysiological *mechanisms* i.e. "entities and activities organized in such a way that they are responsible for the phenomenon" (Illari and Williamson, 2012). Whether at the computational, algorithmic or implementation level (Marr, 1982), mechanistic functional descriptions may help determining when the functional space can be endowed with a given representation, e.g. a network representation, how reproducible it is, and when this representation ceases to be appropriate, and may ultimately have an impact on the ability to predict and act upon brain activity.

**APPENDIX**

- A *topology* on a set $X$ is a collection $\mathscr{T}$ of subsets of $X$, called open sets, satisfying a set of axiomatic properties. The pair $(X, \mathscr{T})$ constitutes a *topological space*.
- A *Hausdorff space* is a topological space with a *separation* property, i.e. any two points can be separated by disjoint open sets.
- The *quotient space* $X/\sim$ of a topological space $X$ and an equivalence relation $\sim$ on $X$ is the set of equivalence classes of points in $X$ under $\sim$ together with the quotient topology. The *quotient topology* consists of all sets with an open preimage under the canonical projection which maps each element to its equivalence class.
- A *topological dynamical system* is a topological space, together with a semigroup of continuous transformations of that space.
- An *n-dimensional manifold* is a topological space where each point has a neighbourhood that is homeomorphic to $\mathbb{R}^N$.
- A *topological manifold* is a locally Euclidean Hausdorff space.
- A *paracompact space* is a topological space in which every open cover has a *locally finite open refinement* i.e. every point of the space has a neighbourhood that intersects only finitely many sets in the cover.
- A *cover* of a set $X$ is a collection of sets whose union contains $X$ as a subset.
- A *differentiable manifold* is a topological manifold equipped with an equivalence class of atlases with differentiable transition maps or functions. A differentiable manifold allows defining tangent vectors and *directional derivatives*. If each transition function is a *smooth map*, the atlas is called a *smooth atlas*, and the manifold itself is called smooth.
- A *chart* of a topological space is a homeomorphism from an open subset $U$ of $X$ to an open subset of $\mathbb{R}^N$.
- An *atlas* of a topological space $X$ is a collection of charts on $X$.
- *Transition maps* provide a way of comparing two charts of an atlas.
- A *fibre bundle* is a space that is locally a product space, but may have a different global topological structure. The similarity between a space $E$ (*total space*) and a product space $B \times F$ (where $B$ is the *base* space, and $F$ the *fibre*) is defined using a continuous surjective map $\pi: E \to B$ (*projection of the bundle*) that in small regions of $E$ behaves like a projection $proj_1: B \times F \to B$.
- An atlas defines the structure of a *fibre bundle* if the transition maps between its charts preserve *local trivializations*. In a fibre bundle, transition maps represent changes in the parameterization of a fibre.
- A *vector bundle* over a space $X$ (e.g. a topological space, a manifold, etc.) is a particular fibre bundle the fibre of which has vector field structure. A vector bundle results from the association of every point $x$ of the space $X$ with a vector space $V(x)$ so that the resulting vector spaces form a space of the same kind as $X$. A vector bundle is in essence a family of vector spaces parameterized by another space X.
- A *fibre bundle section* is a continuous right inverse of the projection function $\pi$ which assigns in a continuous manner a value of the fibre from the attached space to every point of the base. In general, fibre bundles do not have global sections, and sections are defined locally. Local sections play the same role as vector fields in an open subset of a topological space.
- Local sections form a *sheaf* over the base, called the *sheaf of the sections* of a space.
- A *sheaf* is a topological object through which locally defined data attached to the open sets of a topological space can be systematically tracked. Obstructions to the existence of a section can sometimes be measured by a *cohomology class*.
- Two flows $f: A \to A$ and $g: B \to B$ are *topologically equivalent* if there exists a homeomorphism $h: A \to B$ that maps the orbits of $f$ onto the orbits of $g$ and preserves the direction of time.
- Two flows $f: A \to A$ and $g: B \to B$ are *topologically conjugate* if there is a homeomorphism $h: A \to B$ such that $g \circ h = h \circ f$
- Generally speaking, interpreting data means transitioning from a *measurable* space $(X, \Sigma)$ to *measure* space $(X, \Sigma, \mu)$, where $X$ is a set, $\Sigma$ a $\sigma$-*algebra* on $X$, and $\mu: \Sigma \to \mathbb{R}^N$ a finite measure on the $\sigma$-algebra. A *probability space* is a particular measure space where $\mu$ is a *probability measure*.
- A $\sigma$-*algebra* on a set $X$ is a collection $\Sigma$ of subsets of $X$ that includes the null set $\emptyset$ and the entire space $X$, and is closed under complement and countable unions.
- From a measure-theoretical viewpoint, a *dynamical system* is a measure-preserving transformation of a $\sigma$-algebra, i.e. a quadruple $(X, \Sigma, \mu, \tau)$, where $\tau: X \to X$ is a *measure-preserving* transformation of $X$. A map $\tau: X \to X$ is a *measure-preserving* transformation of $X$ if it is $\Sigma$-measurable, and is measure-preserving. $\tau$ is $\Sigma$-measurable if and only if for every $\sigma \in \Sigma$, $\tau^{-1}\sigma \in \Sigma$ and preserves the measure if and only if for every $\sigma \in \Sigma, \mu(\tau^{-1}\sigma) = \mu(\sigma)$.

**Acknowledgements** - The author acknowledges financial support from the program *Accueil de Talents* of the Métropole Européenne de Lille.